\documentclass[12pt]{article}

\usepackage{amsmath,graphicx}
\usepackage{multirow}
\usepackage{amsfonts}
\usepackage{amssymb}
\usepackage{amscd}
\usepackage{cite}
\usepackage{amsmath}
\usepackage{bbm}
\usepackage{color}

\def\hybrid{\topmargin -20pt    \oddsidemargin 0pt
        \headheight 0pt \headsep 0pt
        \textwidth 6.25in       
        \textheight 9.5in       
        \marginparwidth .875in
        \parskip 5pt plus 1pt   \jot = 1.5ex}

\hybrid

\newcommand{\be}{\begin{equation}}
\newcommand{\ee}{\end{equation}}

\newcommand{\bea}{\begin{eqnarray}}
\newcommand{\eea}{\end{eqnarray}}

\renewcommand{\Re}{\operatorname{Re}}
\renewcommand{\Im}{\operatorname{Im}}

\newcommand\e{\mathrm{e}}
\newcommand\iu{\operatorname{i}}
\newcommand\diff{\mathrm{d}}

\newcommand\nc{{n_{\rm c}}}
\newcommand\nr{{n_{\rm r}}}

\newcommand{\ba}{\begin{eqnarray*}}
\newcommand{\ea}{\end{eqnarray*}}
\newcommand{\ban}{\begin{eqnarray}}
\newcommand{\ean}{\end{eqnarray}}

\newcommand{\cN}{{\cal N}}

\begin{document}
\begin{titlepage}
\begin{center}
\rightline{\small ZMP-HH/12-6}
\rightline{\small IPhT-t12/031}
\vskip 1cm

{\Large \bf
Supersymmetric Vacua in $N=2$ Supergravity}	
\vskip 1.2cm
{\bf Jan Louis$^{a,b}$, Paul Smyth$^{c}$ and Hagen Triendl$^{d}$}

\vskip 0.8cm

$^{a}${\em II. Institut f\"ur Theoretische Physik der Universit\"at Hamburg, Luruper Chaussee 149, D-22761 Hamburg, Germany}
\vskip 0.4cm
$^{b}${\em Zentrum f\"ur Mathematische Physik,
Universit\"at Hamburg,\\
Bundesstrasse 55, D-20146 Hamburg, Germany}
\vskip 0.4cm
$^{c}$ {\em Institut de Th\'eorie des Ph\'enom\`enes Physiques, EPFL\\ CH-1015 Lausanne, Switzerland}
\vskip 0.4cm
$^{d}$ {\em Institut de Physique Th\'eorique, CEA Saclay\\
Orme des Merisiers, F-91191 Gif-sur-Yvette, France\\
}
\vskip 0.8cm

{\tt jan.louis@desy.de, paul.smyth@epfl.ch, hagen.triendl@cea.fr}

\end{center}

\vskip 1cm

\begin{center} {\bf ABSTRACT } \end{center}

\noindent
We use the embedding tensor formalism to analyse maximally symmetric backgrounds of $N=2$ gauged
supergravities which have the full $N=2$ supersymmetry.
We state the condition for $N=2$ vacua and discuss some of their
general properties.
We show that if the gauged isometries leave the $SU(2)$ R-symmetry
invariant, then the $N=2$ vacuum must be Minkowski.
This implies that there are no AdS backgrounds with eight unbroken supercharges in the effective $N=2$ supergravity of six-dimensional $SU(3)\times SU(3)$ structure compactifications of type II string theory and M-theory.
Combined with
previous results on $N=1$ vacua, we show that there exist $N=2$ supergravities
with a given set of gauged Abelian isometries that have both $N=2$ and $N=1$
vacua. We also argue
that an analogue of our analysis holds in five and six spacetime
dimensions.

\vfill

April 2012

\end{titlepage}

\section{Introduction}

The analysis of Minkowski and Anti-de Sitter (AdS) supersymmetric
vacua in gauged extended supergravity has received much
attention in recent years. In this paper we consider such maximally-symmetric backgrounds of $N=2$ supergravities in four spacetime dimensions ($d=4$) and their ``cousins''  in $d=5,6$ which also have eight supercharges.

The general conditions for $N=2$ vacua in electrically gauged $N=2$
supergravities, together with a few illustrative examples,
were given recently in \cite{Hristov:2009uj}.
By using the embedding tensor formalism, introduced in
\cite{deWit:2005ub} and applied to $N=2$ gauged supergravity in \cite{deWit:2011gk},\footnote{Related work on tensor fields in $N=2$ supergravity has been performed in \cite{Huebscher:2010ib}.}
we extend the analysis of \cite{Hristov:2009uj} by
allowing for the possibility of electrically and magnetically charged fields in the spectrum.
We derive the conditions for $N=2$ vacua with Abelian and non-Abelian
factors in the gauge group and show that
solutions generically exist.
However, it is not guaranteed that these solutions lie inside
the physical domain of the K\"ahler cone and thus are physically acceptable.
For the special case of hypermultiplets that are gauged with respect to isometries which do not induce an $SU(2)$ R-symmetry rotation, we show that $AdS$ vacua with eight unbroken supercharges are not possible. It is straightforward to extend our analysis to spacetimes with $d=5,6$.

We shall specifically study the class of isometries that are
present in quaternionic-K\"ahler manifolds which are in the image of
the c-map  and appear at the tree
level of type II compactifications in string theory
\cite{Cecotti:1988qn,Ferrara:1989ik}. These manifolds can be viewed as a graded Heisenberg algebra fibred over a special-K\"ahler base. We show that no $N=2$ $AdS$ vacua can occur for gauged isometries in the fibre, which in turn implies that there are no $AdS$ vacua in the low-energy effective $N=2$ action of six-dimensional $SU(3)\times SU(3)$-structure compactifications of type II string theory and M-theory that preserve eight supercharges. This means in particular that $SU(3)\times SU(3)$-structure backgrounds with four- and five-dimensional $N=2$ AdS vacua as found in \cite{Gauntlett:2004hh,Petrini:2009ur,Lust:2009mb}  do not have any description in terms of $N=2$ gauged supergravity.
The conditions for $N=2$ Minkowski vacua are linear in the fibre coordinates and holomorphic in the coordinates on the special-K\"ahler manifold suggesting that generically a solution exists. However, the K\"ahler cone condition is not automatically satisfied for these solutions.

$N=2$ supergravities
with $N=1$ vacua were first discovered in Refs.~\cite{Ferrara:1995gu,Ferrara:1995xi,Fre:1996js} and later systematically
analysed in \cite{Louis:2009xd,Louis:2010ui,Cortes:2011ut}. It is of interest to determine under what conditions
these supergravities can also admit $N=2$ vacua in their
field space. We again find that the conditions are linear in the fibre coordinates and holomorphic in the special-K\"ahler coordinates, leaving the K\"ahler cone condition as the non-trivial requirement to find a physically acceptable
solution. We give two examples of special-K\"ahler manifolds with cubic prepotential, one of which contains either an $N=1$ or an $N=2$ vacuum inside the K\"ahler cone but never both at the same time. The second example can accommodate both $N=1$ and $N=2$ vacua inside the K\"ahler cone, as long as the charges are chosen appropriately.

This paper is organised as follows. In Section~\ref{sugra} we briefly
introduce $N=2$ gauged supergravity in order to set the stage for the analysis.
In Section~\ref{N2} we record the conditions for vacua with the
full $N=2$ supersymmetry and determine some of their properties.
In Section~\ref{d56} we extend the analysis to supergravities with
eight supercharges in $d=5,6$. In Section~\ref{sec:c-map} we consider
the special case of gauged isometries in the fibre of
quaternionic-K\"ahler manifold which are in the image of the c-map.
Finally, in Section~\ref{sec:N2vsN1} we address the question of
simultaneously having $N=2$ and $N=1$ vacua in the same gauged supergravity.

\section{Gauged supergravity with eight supercharges}\label{sugra}

Let us start with a brief summary of
gauged $N=2$ supergravity in $d=4$.\footnote{For a more comprehensive
  review see, for example,  Ref.~\cite{Andrianopoli:1996cm}.}
Its spectrum consists of a gravitational multiplet, $n_{\rm v}$ vector multiplets and $n_{\rm h}$ hypermultiplets.\footnote{We neglect the possibility of tensor multiplets, as they can be dualised into hypermultiplets (or vector multiplets, if they are massive).} The gravitational multiplet contains the
spacetime metric $g_{\mu \nu}$,
two gravitini $\Psi_{\mu {\cal A}},{\cal A}=1,2 $ and the graviphoton $A^0_\mu$.
Each vector multiplet contains a vector $A^i_\mu$, two gaugini $\lambda^{i\cal A}$ and a complex scalar $t^i$,
where $i=1,\dots,n_{\rm v}$ labels the vector multiplets.\footnote{Strictly speaking, the definition of the graviphoton is $X^I \Im {\cal F}_{IJ} A^J_\mu$, which can be read off from the gravitino variation and depends on the scalar fields in the vector multiplets.}
Finally, a hypermultiplet consists of two hyperini $\zeta_\alpha$ and four scalars $q^u$, where $\alpha =1,\dots, 2n_{\rm h}$ and $u=1,\dots, 4n_{\rm h}$.
The scalar field space is parametrised by $(t^i, q^u)$ and splits into the product
\begin{equation}\label{Fspace}
M = M_{\rm v}\times M_{\rm h} \ .
\end{equation}
The first component $M_{\rm v}$ is a special-K\"ahler manifold of complex
dimension $n_{\rm v}$ spanned by the scalars $t^i$ in the vector multiplets. This implies that the metric obeys
\begin{equation}\label{Kdef}
g_{i\bar \jmath} = \partial_i \partial_{\bar \jmath} K^{\rm v}\ ,
\qquad \textrm{with}\qquad
K^{\rm v}= -\ln \iu \left( \bar X^\Lambda \Omega_{\Lambda \Sigma}
   X^\Sigma \right)\ ,
\end{equation}
where $X^\Lambda = (X^I, {\cal F}_I), I=0,\ldots,n_{\rm  v}$ is a
$2(n_{\rm  v}+1)$-dimensional symplectic vector that
depends holomorphically on the $t^i$.
${\cal F}_I = \partial{\cal F}/\partial X^I $ is the derivative of a
holomorphic prepotential ${\cal F}$ which is homogeneous of degree 2
and $\Omega_{\Lambda \Sigma}$ is the standard symplectic metric.
The physical range of
the coordinates $t^i$ is restricted to the K\"ahler cone defined by
\begin{equation} \label{eq:Kahlercone}
\iu \left( \bar X^\Lambda \Omega_{\Lambda \Sigma} X^\Sigma \right) > 0 \ .
\end{equation}

The second component of the field space
$M_{\rm h}$, spanned by the scalars $q^u$ in the hypermultiplets,
is quaternionic-K\"ahler and of real dimension $4n_{\rm h}$. These
manifolds admit a triplet of almost complex structures $I^x, x=1,2,3$ satisfying
$I^x I^y = - \delta^{xy} {\bf 1} + \epsilon^{xyz} I^z$, with
the metric being Hermitian with respect to all three $I^x$. The
associated two-forms $K^x$
are the field strength of the $SU(2)$ connection $\omega^x$, i.e.\
\begin{equation}\label{Kx}
 K^x = \diff \omega^x + \tfrac12 \epsilon^{xyz} \omega^y \wedge \omega^z \ .
\end{equation}

In gauged supergravities the multiplets can be charged
under a set of  electric and magnetic gauge fields.
The corresponding covariant derivatives of the scalars  read
\begin{equation}\label{Dcov}
 D_\mu q^u = \partial_\mu q^u - A_\mu^\Lambda \Theta_\Lambda^\lambda k^u_\lambda \ , \qquad
D_\mu t^i = \partial_\mu t^i - A_\mu^\Lambda {\hat \Theta}_\Lambda^{\hat \lambda} k^i_{\hat \lambda} \ ,
\end{equation}
where $A_\mu^\Lambda= (A^I_\mu, B_{\mu\,I})$ is a symplectic vector of electric and magnetic gauge fields
and $k^u_\lambda~ (k^i_{\hat \lambda}), \lambda= 1 , \dots, n_{\rm Kh}, (\hat \lambda= 1 , \dots, n_{\rm Kv},)$ are
Killing vectors  on $M_{\rm h} ~(M_{\rm v})$ respectively.
Finally, the charges or group theoretical
representations of the scalars are specified by  the embedding tensors
$\Theta_\Lambda^\lambda, {\hat \Theta}_\Lambda^{\hat \lambda}$.
Note that the $t^i$ transform in the adjoint representation of
the gauge group and thus for any non-Abelian factor
the gauged $k^i_{\hat \lambda}$ have to be non-trivial.
Moreover, if the gauged isometries are non-Abelian, the embedding tensor has to transform covariantly, which is ensured by the quadratic constraint
\begin{equation} \label{eq:quadraticconstraint}
  f^{\hat \lambda}_{\hat \sigma \hat \rho} {\hat \Theta}_\Lambda^{\hat \sigma} {\hat \Theta}_\Sigma^{\hat \rho} + {\hat \Theta}_\Lambda^{\hat \sigma} (k_{\hat \lambda})^{\Gamma}_{\Sigma} {\hat \Theta}_\Gamma^{\hat \lambda} = 0 \ .
\end{equation}
Here $(k_{\hat \lambda})^{\Gamma}_{\Sigma}$
is the symplectic transformation induced by the Killing vector $k^i_{\hat \lambda}$ via
\begin{equation}\label{eq:Killingsymplectic}
 k^i_{\hat \lambda} \partial_i X^\Lambda = (k_{\hat \lambda})^{\Lambda}_{\Sigma} X^\Sigma \ ,
\end{equation}
and $f^{\hat \lambda}_{\hat \sigma \hat \rho}$ are the structure constants
\begin{equation}\label{eq:structureconstants}
[k_{\hat \sigma},k_{\hat \rho}] = f^{\hat \lambda}_{\hat \sigma \hat \rho} k_{\hat \lambda} \ .
\end{equation}
Note that both $(k_{\hat \lambda})^{\Gamma}_{\Sigma}$ and $f^{\hat \lambda}_{\hat \sigma \hat \rho}$ are independent of the coordinates $t^i$.

The gauging of isometries requires additional terms in the
supersymmetry variations. Since we are looking for maximally symmetric
backgrounds it is sufficient to focus on the
scalar parts of the fermionic supersymmetry variations given by
\begin{equation}\label{eq:susyvar}\begin{aligned}
 \delta_\epsilon \Psi_{\mu {\cal A}} &= D_\mu \epsilon^*_{\cal A} - S_{\cal AB} \gamma_\mu \epsilon^{\cal B} + \ldots \ ,\\
\delta_\epsilon \lambda^{i {\cal A}} &= W^{i{\cal AB}}\epsilon_{\cal B}+\ldots \ ,\\
\delta_\epsilon \zeta_{\alpha} &= N_\alpha^{\cal A} \epsilon_{\cal A}+\ldots \ ,
\end{aligned}\end{equation}
where $\epsilon^{\cal A}$ are the supersymmetry parameters
and
\begin{equation}\label{eq:matrices}\begin{aligned}
S_{\cal AB} =&\, \tfrac{1}{2} \e^{K^{\rm v}/2} X^\Lambda \Theta_\Lambda^{\lambda} P_{\lambda}^x
(\sigma^x)_{\cal AB} \ , \\
W^{i{\cal AB}}
=&\, \mathrm{i} \e^{K^{\rm v}/2} g^{i\bar \jmath}\,
(\nabla_{\bar \jmath}\bar X^\Lambda) \Theta_\Lambda^{\lambda} P_{\lambda}^x (\sigma^x)^{\cal AB} + \e^{K^{\rm v}/2} \epsilon^{\cal AB} \bar X^\Lambda {\hat \Theta}_\Lambda^{\hat \lambda}  k^i_{\hat \lambda}
\ , \\
N_\alpha^{\cal A}
=&\, 2 \e^{K^{\rm v}/2} \bar X^\Lambda \Theta_\Lambda^{\lambda} {\cal U}^{\cal A}_{\alpha u} k^u_{\lambda}
\ .
\end{aligned}\end{equation}
Here ${\cal U}^{\cal A \alpha}$ are the vielbein one-forms on $M_{\rm
  h}$, the $(\sigma^x)^{\cal A}_{\cal B}$ are the Pauli matrices,
and
$\nabla_{i}X^\Lambda := \partial_i X^\Lambda + (\partial_i K^{\rm v})X^\Lambda$.
Finally, $P_{\lambda}^x$ are the Killing prepotentials defined by
\begin{equation}\label{eq:Pdef}
-2 k^u_\lambda\,K_{uv}^x  =   \nabla_v P_\lambda^x\ ,
\end{equation}
where $\nabla_v$ is the $SU(2)$-covariant derivative and the two-forms
$K^x$ are defined in \eqref{Kx}.
The matrices given in \eqref{eq:matrices} also
determine the scalar
potential $V$ in the Lagrangian
\begin{equation}
 V = -6 S_{\cal AB} \bar S^{\cal AB} + \tfrac12 g_{i \bar \jmath} W^{i{\cal AB}} W^{\bar \jmath}_{\cal AB} + N_\alpha^{\cal A} N^\alpha_{\cal A} \ .
\end{equation}

To conclude, a gauged supergravity is specified by the spectrum of
vector- and hypermultiplets, their respective field spaces and the
embedding tensor which determines the charged directions in field space.

\section{Vacua with $N=2$ supersymmetry}\label{N2}

We shall now give the conditions for vacua which have
the full $N=2$ supersymmetry.
This requires that all fermionic supersymmetry variations \eqref{eq:susyvar} vanish, which, for a maximally symmetric spacetime, translates into the conditions
\begin{equation}\label{eq:susy}
S_{\cal AB} \epsilon^{\cal B} = \tfrac12 \mu \epsilon^*_{\cal A}\ ,\qquad
 W^{iAB} = 0 \ , \qquad N^{\alpha A} = 0 \ ,
\end{equation}
where $\Lambda = -3 |\mu|^2$ is the cosmological constant of the $N=2$ vacuum. These conditions have been discussed before for electric gaugings in \cite{Hristov:2009uj}.

Let us start by analysing the second condition in \eqref{eq:susy}.
Since  $(\sigma^x)^{\cal AB}$ and $ \epsilon^{\cal AB}$ are linearly
independent, this condition together with the definition \eqref{eq:matrices} implies~\cite{Hristov:2009uj}
\begin{eqnarray}\label{eq:gaugino_one}
(\nabla_{i}  X^\Lambda)\, \Theta_\Lambda^{\lambda} P_{\lambda}^x & =& 0 \ , \\
\bar X^\Lambda {\hat \Theta}_\Lambda^{\hat \lambda}  k^i_{\hat \lambda} & =& 0 \ .\label{eq:gaugino_two}
\end{eqnarray}
Equation \eqref{eq:gaugino_two}  only depends on the $t^i$
and has a trivial solution $ k^i_{\hat \lambda} =0$ with the property
that any non-Abelian factor of the gauge group is
unbroken in the vacuum. If, on the other hand,
the background has $k^i_{\hat \lambda} \neq 0$, the gauge group is
spontaneously broken and \eqref{eq:gaugino_two}
can only be fulfilled by tuning some of the $t^i$'s appropriately.
Contracting \eqref{eq:gaugino_two} with $\partial_i X^\Sigma$ and using
\eqref{eq:Killingsymplectic} yields
\begin{equation}\label{int2}
\bar X^\Lambda {\hat \Theta}_\Lambda^{\hat \lambda}  (k_{\hat \lambda})^\Sigma_\Gamma X^\Gamma = 0 \ ,
\end{equation}
which, upon further multiplication with ${\hat \Theta}^{\hat \rho}_\Sigma$
and use of \eqref{eq:quadraticconstraint}, results in
\begin{equation}
\iu \bar X^\Lambda ({\hat \Theta}_\Lambda^{\hat \lambda} f^{\hat \rho}_{\hat \lambda \hat \sigma} {\hat \Theta}_\Gamma^{\hat \sigma})  X^\Gamma = 0 \ .
\end{equation}
This gives a number of real quadratic equations for $X^\Lambda$, which fix
$\nr=\operatorname{rk}(T(t, \bar t))$ real degrees of freedom at some point $t^i$, where we defined the $n_{\rm Kv} \times (4n_{\rm v} +4)$-matrix
\begin{equation}
T^{\hat \rho}_{\hat \Lambda}(t, \bar t) =\left(-{\hat \Theta}_\Lambda^{\hat \lambda} f^{\hat \rho}_{\hat \lambda \hat \sigma} {\hat \Theta}_\Gamma^{\hat \sigma}  \Im(X^\Gamma(t)), {\hat \Theta}_\Lambda^{\hat \lambda} f^{\hat \rho}_{\hat \lambda \hat \sigma} {\hat \Theta}_\Gamma^{\hat \sigma}  \Re(X^\Gamma(t))\right) \ , \qquad \hat \Lambda = 1 , \dots , 4n_{\rm v} +4 \ .
\end{equation}
As a consequence $\nr$ gauge bosons
become massive by ``eating'' $\nr$ real scalar degrees of freedom
leaving $\nr$ massive short BPS vector multiplets.\footnote{Note that for $k^i_{\hat \lambda}=0$ we have $T^{\hat \rho}_{\hat \Lambda}=0$ and therefore $\nr=0$ so that the gauge group remains unbroken.}

Before we analyse
\eqref{eq:gaugino_one} let us turn to the third condition in \eqref{eq:susy}.
Since the vielbein on the quaternionic-K\"ahler manifold is invertible
we infer from \eqref{eq:matrices} that  $N^{\alpha A} = 0$ implies
\begin{equation}\label{eq:hyperino}
 X^\Lambda \Theta_\Lambda^\lambda k_\lambda^u = 0 \ ,
\end{equation}
which is similar to \eqref{eq:gaugino_two} but now couples the
vector- and hypermultiplet sector. Furthermore, in contrast to \eqref{eq:gaugino_two}, equations \eqref{eq:hyperino} are holomorphic conditions on the $t^i$.
As before there is the trivial solution $k_\lambda^u = 0$ but
 \eqref{eq:hyperino}
can also be satisfied by  tuning further vector scalars $t^i$ appropriately.
More precisely,
from the Killing vectors $k_\lambda$ that are non-zero
at the
vacuum locus $\nc = \operatorname{rk}(\Theta_\Lambda^\lambda k_\lambda^u)$ holomorphic conditions arise for the
vector multiplet scalars $t^i$ which in turn imply that there are
$\nc$  further massive gauge boson.\footnote{Note that
electric gaugings give rise to linear equations, while magnetic gaugings are non-linear in the standard coordinates on $M_{\rm v}$.}
As we shall see shortly, these
massive gauge bosons reside in long non-BPS vector multiplets.
Note that the combined conditions following from \eqref{eq:gaugino_two} and
\eqref{eq:hyperino} have to be compatible and solvable by tuning at most
$n_{\rm v}$ complex scalars.

Now let us turn to \eqref{eq:gaugino_one} which
can be nicely combined with the first equation in \eqref{eq:susy}.
Noting that the matrix $(X^I, \nabla_i X^I)$ is invertible in special geometry
we can rewrite the two conditions together as \cite{Louis:2009xd}
\begin{equation}\label{int}
( \Theta_I^{\, \lambda} - {\cal F}_{IJ} \Theta^{J \lambda})P^x_\lambda
= - \e^{- K^{\rm v}/2} (\partial_I K^{\rm v})\, \hat \mu\, a^x  \ ,
\end{equation}
where $a^x$ is an arbitrary real vector on $S^2$ and $\hat \mu$ is
related to $\mu$ by a phase. From the definition of the K\"ahler potential \eqref{Kdef} we have $X^I \partial_I K^{\rm v} = 1$ and $(\partial_i X^I) \partial_I K^{\rm v} = 0$, which means that the right-hand side in \eqref{int} gives only a contribution to the gauging of the graviphoton $X^I \Im {\cal F}_{IJ} A^J_\mu$.\footnote{This explicit expression for the graviphoton is found from its appearance in the gravitino variation.} The non-vanishing prepotential of this gauging therefore determines the cosmological constant, while the prepotentials of all other gaugings should vanish in an $N=2$ vacuum.
We can easily solve \eqref{int} for the prepotentials. Since $\Im {\cal F}_{IJ}$ is required by special geometry to be invertible,
\eqref{int} is equivalent to
\begin{equation}\label{eq:gaugino}
\Theta_\Lambda^{\, \lambda}P^x_\lambda = - \tfrac12 \e^{K^{\rm v}/2} \Omega_{\Lambda \Sigma} \Im (\hat \mu \bar X^\Sigma)\, a^x  \ ,
\end{equation}
where we used \eqref{Kdef} and
$\Theta_\Lambda^{\, \lambda} =(\Theta_I^{\, \lambda}, - \Theta^{J \lambda})$.

In general \eqref{eq:gaugino} corresponds to
$3 \nc$ real conditions for the
hypermultiplet scalars which in turn become massive.
As we observed above $\nc$
gauge bosons also become massive by each eating the forth
real scalar field of a hypermultiplet.  We thus see that
the Higgs mechanism
leads to a long massive vector multiplet which contains altogether
five massive scalars -- three from hypermultiplets and two from vector multiplets. For $N=2$ Minkoswki vacua those scalar fields which do not participate in the Higgs mechanism are flat directions of the vacuum and thus define its $N=2$ moduli space. For $N=2$ AdS vacua both \eqref{eq:hyperino} and \eqref{eq:gaugino} generate further scalar masses so that the actual moduli space can be much smaller.
Note that we need $n_{\rm h} \ge \nc$ in order to have an $N=2$ vacuum.

Let us now consider the special case of isometries $k_\lambda$ which
do not induce an $SU(2)$ R-symmetry rotation on the fermions,
i.e.\ isometries of $M_{\rm h}$ whose
Lie derivative on the $SU(2)$ connection vanishes
\begin{equation}\label{eq:Liecon}
 {\cal L}_k \omega^x = 0 \ .
\end{equation}
For such isometries the Killing prepotentials
are given in terms of the $SU(2)$ connection by \cite{Michelson:1996pn}
\begin{equation} \label{eq:prepcon}
 P^x = \omega^x(k) \ .
\end{equation}
Inserted into $S_{\cal AB}$ the hyperino condition \eqref{eq:hyperino} implies
\begin{equation}\label{eq:Szero}
S_{\cal AB} \sim X^\Lambda \Theta_\Lambda^\lambda P^x_\lambda (\sigma^x)^{AB} =  \omega^x(X^\Lambda \Theta_\Lambda^\lambda k_\lambda) (\sigma^x)_{\cal AB} = 0 \ .
\end{equation}
From Eq.~\eqref{eq:susy} we then infer that
the cosmological constant must vanish and all $N=2$ vacua in such
theories are necessarily Minkowski.
It can be easily checked that the isometries in the fibre of
quaternionic-K\"ahler manifolds which are in the image of the c-map, and which we discuss in more detail in Section~\ref{sec:c-map}, have this property \cite{Michelson:1996pn,Louis:2009xd}. Note, however, that there are also examples where \eqref{eq:Liecon} is not fulfilled \cite{Gauntlett:2009zw,Hristov:2009uj}.

Before we proceed, let us address the issue of the $SU(2)$-covariance of our result. Both \eqref{eq:Liecon} and \eqref{eq:prepcon} do not transform covariantly under local $SU(2)$ rotations and therefore one might worry that they only hold for a particular choice of coordinates.\footnote{We thank the referee and S.\ Vandoren for drawing our attention to this subtlety.} Indeed, the Killing prepotentials can be written more generally as \cite{D'Auria:1990fj}
\begin{equation}
 P_\lambda^x = \omega^x(k_\lambda) +W_\lambda^x \ ,
\end{equation}
where $W_\lambda^x$ is the so-called compensator field that makes the right-hand side transform non-trivially as an $SU(2)$ vector and that is defined via
\begin{equation} \label{eq:LieKcomp}
 {\cal L}_{k_\lambda} K^x = \epsilon^{xyz} K^y W_\lambda^z \ .
\end{equation}
As a consequence of \eqref{eq:Liecon} the left-hand side of this equation vanishes and the compensator field vanishes in this particular $SU(2)$ frame.
However, in the $N=2$ locus \eqref{eq:hyperino} implies that
\begin{equation}
X^\Lambda \Theta_\Lambda^\lambda P^x_\lambda\Big|_{N=2}  =  X^\Lambda \Theta_\Lambda^\lambda W^x_\lambda\Big|_{N=2}  \ ,
\end{equation}
where each side transforms as an $SU(2)$ vector.
This means that
\begin{equation}\label{eq:LieKvac}
X^\Lambda \Theta_\Lambda^\lambda {\cal L}_{k_\lambda} K^x \Big|_{N=2} = 0 \ ,
\end{equation}
is an $SU(2)$-covariant condition (see also appendix A.3 of \cite{Louis:2009xd} for similar manipulations), which follows from the non-covariant equation \eqref{eq:Liecon}. Furthermore, the condition \eqref{eq:LieKvac} implies that $S_{\cal AB}$ is vanishing and that the $N=2$ vacuum must be Minkowski.

\section{$N=2$ supergravities in $d=5,6$}\label{d56}

The analysis of the previous section can be repeated in five and six dimensions for supergravities with the same number (eight) of  supercharges.  The hypermultiplet sector is unchanged while
the vector multiplets have only one real scalar in $d=5$ or none at
all in $d=6$. As a result the matrices appearing in fermionic supersymmetry variations \eqref{eq:susyvar} change.

Five-dimensional $N=2$ gauged supergravity has been discussed for
example in \cite{Gunaydin:1999zx,Bergshoeff:2004kh,Zagermann:2001ki} and references therein. Here we will restrict to the case with no tensor multiplets and comment on the more general case later.
The $N=2$ vacua again arise as solutions of \eqref{eq:susy}, with the major difference relative to $d=4$ being that there are no
magnetically charged fields,
as there are no magnetic gauge vectors. In addition,
the scalar matrices previously defined in \eqref{eq:matrices} now read
\begin{equation}\label{eq:matrices5d}\begin{aligned}
S_{\cal AB} =&  h^I \Theta_I^{\lambda} P_{\lambda}^x
(\sigma^x)_{\cal AB} \ , \\
W^{i{\cal AB}}
=& -\tfrac{\sqrt{3}}{\sqrt{2}} g^{ij} \partial_j h^I \Theta_I^{\lambda} P_{\lambda}^x (\sigma^x)^{\cal AB} \ , \\
N_\alpha^{\cal A}
=& \tfrac{\sqrt{6}}{4} h^I \Theta_I^{\lambda} {\cal U}^{\cal A}_{\alpha u} k^u_{\lambda}
\ ,
\end{aligned}\end{equation}
and depend in the vector multiplets only on a set
of real coordinates $h^I$ (instead of the complex coordinates $X^I$) that obey the cubic condition
\begin{equation}\label{5d_hypersurface}
d_{IJK} h^I h^J h^K = 1 \ .
\end{equation}
Analogously to the derivation of \eqref{eq:hyperino}, the hyperino condition $N_\alpha^{\cal A}=0$ leads to
\begin{equation}\label{eq:hyperino5d}
h^I \Theta_I^\lambda k_\lambda^u = 0 \ .
\end{equation}
These are $\nr=\mathrm{rk}(\Theta_I^\lambda k^u_\lambda)$ real equations on the $h^I$ which fix the scalars of $\nr$ vector multiplets.
Furthermore, $(h^I, \partial_j h^I)$ is again an invertible matrix so that, similarly to
\eqref{eq:gaugino}, we can combine the gaugino and gravitino equation into
\begin{equation}\label{eq:gaugino5d}
\Theta_I^{\lambda}P^x_\lambda = d_{IJK} h^J h^K \mu a^x  \ ,
\end{equation}
where  $\mu$ is real and $d_{IJK} h^J h^K$ replaces $\partial_I K$ in \eqref{eq:gaugino} by virtue of the cubic condition \eqref{5d_hypersurface}.
This fixes $3 \nr$ hypermultiplet scalars, consistent with the Higgs mechanism
and we end up with $\nr$ long massive vector multiplets.
Note that, analogously to four dimensions, a supersymmetric AdS vacuum exists only if the Lie derivative on the $SU(2)$ connection is non-zero for at least one of the gauged Killing vectors. The story gets more involved in the presence of tensor multiplets \cite{Gunaydin:1999zx}. However, let us stress that the cosmological constant is only affected by gaugings in the hypermultiplets, and therefore our discussion concerning the existence of supersymmetric AdS vacua still applies.

We now turn to
gauged supergravities with eight supercharges in $d=6$ which
are discussed, for example, in \cite{Nishino:1986dc,Riccioni:2001bg}.
In this case  there are no scalars in the vector multiplet
sector. Moreover, due to chirality of the supergravity no scalar
contributions  arise in the hyperino or gravitino variation, in
contrast to \eqref{eq:susyvar}. From the gaugino variation one finds similarly to \eqref{eq:hyperino} the condition
\begin{equation}
\Theta_i^{\lambda}P^x_\lambda = 0 \ ,
\end{equation}
which again are $3 \operatorname{rk}(\Theta)$ real conditions on the hypermultiplet scalars, as required by the Higgs mechanism. Furthermore, supersymmetric AdS is not a solution, as gaugings do not give a contribution to the cosmological constant.

\section{Gauging the isometries of the c-map}\label{sec:c-map}
A large class of known quaternionic-K\"ahler manifolds are those that lie in
the image of the c-map \cite{Cecotti:1988qn,Ferrara:1989ik}. These
manifolds are fibrations of a graded Heisenberg algebra over a
special-K\"ahler manifold and they are of
interest as the fibre admits a large number of isometries.
Furthermore,
they appear in the low-energy effective action of type II and M-theory
compactifications on six-dimensional $SU(3)\times SU(3)$ structure manifolds
where fluxes, torsion and non-geometric fluxes precisely
gauge these isometries (see e.g. \cite{Polchinski:1995sm,Michelson:1996pn,Gurrieri:2002wz,D'Auria:2004tr,Grana:2005ny,Grana:2006hr,Cassani:2007pq}).
Therefore the vacua in these
gauged supergravities coincide with the vacua for $SU(3)\times SU(3)$-structure compactifications of type II and M-theory to four and five dimensions, respectively.

Let us denote the $(n_{\rm h}-1)$ complex
coordinates of the special-K\"ahler base space by~$z^a$,
the analogue of the holomorphic symplectic vector $X^\Lambda$ by
$Z^{\tilde \Lambda} = (Z^A, {\cal G}_A)$ and the corresponding
K\"ahler potential by $K^{\rm h}$. The c-map adds an additional $(2n_{\rm h}+2)$
real fibre coordinates $(\phi, \tilde \phi,\xi^{\tilde \Lambda})$ where
$\xi^{\tilde \Lambda} = (\xi^A, \tilde \xi_A)$ is a $2n_{\rm h}$-dimensional
symplectic vector.\footnote{For more details see, for example,
  \cite{Cecotti:1988qn,Ferrara:1989ik,Louis:2009xd}.} The isometries
of the fibre are generated by the Killing vectors\footnote{We neglect the Killing vector in the $\phi$ direction, as this isometry is  broken in string compactifications by one-loop corrections \cite{RoblesLlana:2006ez}.}
\begin{equation}\label{eq:Killing}
\begin{aligned}
k_{\tilde \phi}  \ &=\ - 2\, \frac{\partial}{\partial {\tilde \phi}} \ , \qquad
 k_{\tilde \Lambda}  \, &= \ \frac{\partial}{\partial \xi^{\tilde \Lambda} } + \Omega_{\tilde \Lambda \tilde \Sigma} \xi^{\tilde \Sigma} \, \frac{\partial}{\partial {\tilde \phi}} \ ,
\end{aligned}
\end{equation}
which form a graded Heisenberg algebra with the only non-trivial commutator being
\begin{equation}
[  k_{\tilde \Lambda} ,  k_{\tilde \Sigma} ] = \Omega_{\tilde \Lambda \tilde \Sigma} k_{\tilde \phi} \ .
\end{equation}

Since these Killing vectors  are everywhere linearly independent,
eq.~\eqref{eq:hyperino} simplifies to
\begin{equation} \label{eq:condV}
 X^\Lambda \Theta_\Lambda^{\tilde \Lambda}= 0 \ , \qquad X^\Lambda \Theta_\Lambda^{\tilde \phi}= 0  \ .
\end{equation}
This gives $\nc=\operatorname{rk}(\Theta)$ holomorphic conditions on
$M_{\rm v}$, giving a mass to $\nc$ vector multiplets in the Higgs mechanism. Furthermore, Eq.~\ref{eq:condV} defines the $N=2$ vector moduli space of the vacuum.

Let us continue with the constraints in the hypermultiplet sector.
The isometries generated by \eqref{eq:Killing} fulfil
\eqref{eq:Liecon} and therefore the $N=2$ vacuum is necessarily Minkowski.
Inserting \eqref{eq:prepcon} into \eqref{eq:gaugino} we arrive at
\begin{equation}\label{eq:condN2}
\Theta_\Lambda^{\tilde \Lambda} \omega^x (k_{\tilde \Lambda}) = 0 \ ,\qquad \Theta_\Lambda^{\tilde \phi} \omega^x (k_{\tilde \phi}) = 0\ .
\end{equation}
The explicit form of the $SU(2)$ connection is given by
\cite{Ferrara:1989ik,Louis:2009xd}
\begin{equation}\label{eq:quat_connection}
\begin{aligned}
&\omega^1 - \iu \omega^2\ =\  2 \e^{K^{\rm h}/2+\phi}Z^A(\diff \tilde\xi_A - {\cal F}_{AB} \diff \xi^B)  \ , \\
&
\omega^3\  =\  \tfrac{1}{2} \e^{2\phi} (\diff \tilde \phi +\tilde\xi_A \diff \xi^A-\xi^A \diff \tilde \xi_A  )
- \iu \e^{K^{\rm h}} \left(Z^A (\Im  \mathcal G_{AB})\diff \bar Z^B - \bar Z^A (\Im  \mathcal G_{AB}) \diff Z^B \right) \ .
 \end{aligned}
\end{equation}
Inserted into \eqref{eq:condN2} yields
\begin{align}
\label{eq:condH1}
\Theta^{\tilde \Lambda}_{\Lambda } \Omega_{\tilde \Lambda\tilde \Sigma} Z^{\tilde \Sigma} = 0 \ , \\
\label{eq:condH2}
\Theta^{\tilde \Lambda}_{\Lambda }\Omega_{\tilde \Lambda\tilde \Sigma} \xi^{\tilde \Sigma} = \Theta_{\Lambda \tilde \phi} \ .
\end{align}
The first equation is completely analogous to \eqref{eq:condV} and
gives
$\nc$ holomorphic conditions on the
special-K\"ahler base of  $M_{\rm h}$. The second equation leads to
$\nc$ real conditions on the fibre of $M_{\rm h}$.
The other $\nc$ fibre scalars are eaten by the gauge
vectors so that altogether there are $\nc$
long massive vector multiplets leaving $n_{\rm v} - \nc$
vector and $n_{\rm h}-\nc$ hypermultiplets unfixed and massless.

Note that \eqref{eq:condV} and \eqref{eq:condH1}
are holomorphic equations of the special-K\"ahler coordinates and \eqref{eq:condH2} gives real, linear equations for the fibre.
Therefore they are generically solvable
but it is not automatic that the solution
lies inside the K\"ahler cones for both the $X^I$ and $Z^A$ (cf.\ \eqref{eq:Kahlercone}). We will see this feature
more explicitly in the next section when we discuss some examples.

\section{$N=2$ and $N=1$ vacua in the same gauged supergravity}
\label{sec:N2vsN1}
In Ref.~\cite{Louis:2009xd} the issue of spontaneous $N=2\to N=1$ supersymmetry breaking was considered and the possible $N=1$ vacua of $N=2$ supergravities were classified. It is of interest to determine under which conditions
a given gauged supergravity can have
simultaneously $N=2$ and $N=1$ vacua in its field space.\footnote{We thank Z. Komargodski for a remark which inspired the following analysis.}
Supersymmetry then implies that both vacua  are completely stable \cite{Weinberg:1982id,Cvetic:1992st}. In the following we derive these
conditions and give two explicit examples.
As we will see,
they are separated in scalar field space
and can lie in the same or in different chambers of the K\"ahler cone.

We will concentrate in the following on supergravities that are in the image of the c-map.
For this class the $N=1$ Minkowski solutions of \cite{Louis:2009xd} can be
stated in terms of the embedding tensor as
\begin{equation} \label{eq:MinkowskiN1}
\begin{aligned}
\Theta_\Lambda^{\ \tilde \Lambda} &= \Re \big(  \bar{C}_\Lambda D^{ \tilde \Lambda}  \big) \ , \qquad
\Theta_\Lambda^{\ \tilde \phi} &= \Re \big(  \bar{C}_\Lambda \hat D  \big) \ ,
\end{aligned}
\end{equation}
where the solution is parametrised by two complex lightlike vectors
$C_\Lambda$ and $D^{ \tilde \Lambda}$ satisfying
\begin{equation}\label{eq:lightlike}
\bar C_\Lambda \Omega^{\Lambda \Sigma} C_\Sigma = 0 \ , \qquad \bar D^{ \tilde \Lambda} \Omega_{\tilde \Lambda \tilde \Sigma} D^{ \tilde \Sigma} = 0 \ ,
\end{equation}
and
\begin{equation}\label{eq:cmapMinkN1}
 {C}^J{\cal F}_{JI} (t_{N=1}) = C_I \ , \qquad {D}^B{\cal G}_{BA} (z_{N=1}) = D_A \ , \qquad {D}^{ \tilde \Lambda}\, \Omega_{\tilde \Lambda \tilde \Sigma}\, \xi_{N=1}^{ \tilde \Sigma} = \hat D \ .
\end{equation}
$\hat D$ is a constant and the last equation
fixes two of the scalars $\xi^{ \tilde \Sigma}$.
The first two equations in \eqref{eq:cmapMinkN1}
generically fix all scalars $t^i$ and $z^a$,
but for special theories there can be a moduli space
spanned by $t_{N=1}$ and $z_{N=1}$, respectively
\cite{Cortes:2011ut}.
The structure of the embedding tensor given in \eqref{eq:MinkowskiN1}
defines the gauged supergravity and the conditions
\eqref{eq:lightlike} and \eqref{eq:cmapMinkN1} ensure that it has $N=1$
vacua. Let us now consider under what conditions
these supergravities can also have $N=2$ vacua.

Clearly, the embedding tensor in \eqref{eq:MinkowskiN1} has just rank
two, so that generically there should also exist an $N=2$ vacuum.
Inserting the $\cN=1$ solutions \eqref{eq:MinkowskiN1} into \eqref{eq:condV}, \eqref{eq:condH1} and \eqref{eq:condH2} we find the $N=2$ condition to be
\begin{equation}\label{eq:cmapMinkN2pre}\begin{aligned}
X^\Lambda_{N=2}\, \Re \big(  \bar{C}_\Lambda D^{ \tilde \Lambda}  \big) & = 0 \ , \qquad
X^\Lambda_{N=2}\, \Re \big(  \bar{C}_\Lambda  \hat D  \big) = 0 \ ,\\
 \Re \big(  \bar{C}_\Lambda D^{ \tilde \Lambda}  \big) \, \Omega_{\tilde \Lambda \tilde \Sigma}\, Z^{\tilde \Sigma}_{N=2} & = 0 \ , \qquad
  \Re \big(  \bar{C}_\Lambda D^{ \tilde \Lambda}  \big) \, \Omega_{\tilde \Lambda \tilde \Sigma}\, \xi_{N=2}^{ \tilde \Sigma} =  \Re \big(  \bar{C}_\Lambda  \hat D  \big) \ ,
\end{aligned} \end{equation}
where the subscript $N=2$ indicates that we evaluate the quantity in
the $N=2$ vacuum.
Using the fact that \eqref{eq:cmapMinkN1} holds at some point in
scalar field space and that $\Im F_{IJ}$ and $\Im G_{AB}$ are invertible,
it follows that neither of the complex vectors $C_\Lambda$ and $D^{ \tilde \Lambda}$ are
proportional to a real vector.
Therefore, the
most general solution of \eqref{eq:cmapMinkN2pre} is
\begin{equation}\label{eq:cmapMinkN2} \begin{aligned}
X^\Lambda_{N=2} \bar{C}_\Lambda & = X^\Lambda_{N=2} {C}_\Lambda = 0 \ , \quad D^{ \tilde \Lambda}\,  \Omega_{\tilde \Lambda \tilde \Sigma}\, Z^{\tilde \Sigma}_{N=2} = \bar D^{ \tilde \Lambda} \, \Omega_{\tilde \Lambda \tilde \Sigma}\, Z^{\tilde \Sigma}_{N=2}  =0 \ , \quad
D^{ \tilde \Lambda}\, \Omega_{\tilde \Lambda \tilde \Sigma}\, \xi_{N=2}^{ \tilde \Sigma} = \hat D \ .
\end{aligned} \end{equation}
We see that the condition on the fibre coordinates $\xi^{ \tilde
  \Sigma}$ is the same for $N=1$ and $N=2$ vacua while the conditions
on the scalars $t^i$ and $z^a$ are less restrictive for $N=2$ vacua.
Generically, two
complex $t^i$ and two complex $z^a$ are fixed by
\eqref{eq:cmapMinkN2}. Therefore, gauged supergravities
which admit an $N=1$ vacuum could easily also have
an $N=2$ vacuum. However, it is
not obvious that both vacua lie within the same K\"ahler cone where
\eqref{eq:Kahlercone} holds.

Before we discuss examples where both types of 
vacua are realised, let us discuss their positions in field space.
On the one hand one expects that 
different vacua 
should not intersect in field space. On the other hand one easily imagines
a point in field space which could fulfil both the $N=2$ and $N=1$ conditions \eqref{eq:cmapMinkN2} and \eqref{eq:cmapMinkN1} simultaneously.
However, the K\"ahler cone condition \eqref{eq:Kahlercone}
ensures that $N=1$ and $N=2$ vacua are always separated in field space.
To see this we combine \eqref{eq:cmapMinkN1} and \eqref{eq:cmapMinkN2} 
to arrive at
\begin{equation}\label{eq:XC_orth}
 \bar X^I (\Im {\cal F})_{IJ} C^J = 0 \ ,
\end{equation}
while \eqref{eq:cmapMinkN1} implies
\begin{equation}\label{eq:C_null}
 \bar C^I (\Im {\cal F})_{IJ} C^J = 0 \ .
\end{equation}
Eq.~\eqref{eq:C_null} states that $C^I$ is lightlike while \eqref{eq:XC_orth} means that $C^I$ and $X^I$ are orthogonal to each other.  In the K\"ahler cone defined by \eqref{eq:Kahlercone}, $X^I$ is timelike, contradicting one of these two statements.
Therefore, both conditions cannot be fulfilled simultaneously as long as
\eqref{eq:Kahlercone} holds. Hence, $N=1$ and $N=2$ vacua can only coincide
outside the physical region of the K\"ahler cone. Of course, the same reasoning also holds for the special-K\"ahler base space in the hypermultiplet sector.

We shall now consider the STU model as a first example, where the scalar manifolds are given by
\begin{equation}
M_{\rm v} = \left( \frac{Sl(2,\mathbb{R})}{SO(2)} \right)^3 \ , \qquad M_{\rm h} = \frac{SO(4,4)}{SO(4)^2} \ .
\end{equation}
This means that both the special-K\"ahler manifold $M_{\rm v}$ for the vector multiplets as well as the special-K\"ahler base underlying the  quaternionic-K\"ahler manifold $M_{\rm h}$ are described by the holomorphic prepotential
\begin{equation}\label{eq:STU_prep}
 {\cal F} = \frac{X^S X^T X^U} {X^0} = STU  \ ,
\end{equation}
where we have defined the complex coordinates $S= \frac{X^S} {X^0}$, $T= \frac{X^T} {X^0},\ U= \frac{X^U} {X^0}$ and chosen $X^0 = 1$. Since the equations \eqref{eq:cmapMinkN1} and \eqref{eq:cmapMinkN2} are identical for both special-K\"ahler manifolds, we will only focus on $M_{\rm v}$ in the following. The discussion for $M_{\rm h}$ is completely analogous.
The K\"ahler potential can be computed from \eqref{eq:STU_prep} and is given by
\begin{equation} \label{eq:ex_Kahlerpot}
 K = - \ln(-\iu (\bar S - S) (\bar T - T)(\bar U - U)) \ ,
\end{equation}
so that the K\"ahler cone condition \eqref{eq:Kahlercone} reads
\begin{equation}\label{eq:STU_Kahlercone}
 \Im S \Im T \Im U > 0 \ .
\end{equation}
This gives various domains where either all imaginary parts are positive or two imaginary parts are negative and the third one is positive.
In \cite{Cortes:2011ut} we already discussed the $N=1$ vacuum of this model. In order to find a vacuum inside the K\"ahler cone, we choose
\begin{equation}
C_S = \frac{C^T C^U}{C^0} \ , \qquad C_T = \frac{C^S C^U}{C^0} \ , \qquad C_U = \frac{C^S C^T}{C^0} \ , \qquad C_0 = - \frac{C^S C^T C^U}{(C^0)^2} \ ,
\end{equation}
with $C^0 \ne 0$.
Furthermore, condition \eqref{eq:lightlike} gives
\begin{equation}\label{eq:ex_lightlike}
 \Im \frac{C^S}{C^0} \Im \frac{C^T}{C^0}  \Im \frac{C^U}{C^0} = 0 \ .
\end{equation}
This means that one of the three imaginary parts, say $\Im \frac{C^U}{C^0}$, must vanish. Then the $N=1$ solution is at \cite{Cortes:2011ut}
\begin{equation}
 S_{N=1} = \frac{C^S}{C^0} \ , \qquad  T_{N=1} = \frac{C^T}{C^0} \ ,
\end{equation}
with $U$ arbitrary. On the other hand,
from \eqref{eq:cmapMinkN2} we infer that
a possible $N=2$ vacuum would be located at
\begin{equation}\begin{aligned}
 S_{N=2} = \frac{C^S}{C^0} \ , \quad  T_{N=2} = \frac{\bar C^T}{\bar C^0} \ , \qquad {\rm or\ at}\qquad
 S_{N=2} = \frac{\bar C^S}{\bar C^0} \ , \quad  T_{N=2} = \frac{C^T}{C^0} \ .
\end{aligned}\end{equation}
Checking the K\"ahler cone condition \eqref{eq:STU_Kahlercone} we see that the $N=1$ and $N=2$ solutions can never be both in the same chamber of the K\"ahler cone.
Therefore, we find either an $N=1$ or an $N=2$ vacuum inside the K\"ahler cone, depending on the choice of $C^I$.

Let us now give an example where $N=1$ and $N=2$ vacua do exist
in the same theory and, moreover, in the same domain of the K\"ahler cone.
We consider a supergravity with the field space
\begin{equation}
M_{\rm v} = \frac{Sl(2,\mathbb{R})}{SO(2)} \times \frac{SO(2,n+2)}{SO(2)\times SO(n+2)} \ , \qquad M_{\rm h} = \frac{SO(4,n+4)}{SO(4)\times SO(n+4)} \ .
\end{equation}
$M_{\rm h}$ is in the image of the c-map where the special K\"ahler
base coincides with $M_{\rm v}$ \cite{Cecotti:1988qn}.
Thus the holomorphic prepotential for both spaces is given by
\begin{equation}\label{eq:ex_prep}
 {\cal F} =\frac{X^S (X^T X^U+X^m X^m)} {X^0}= STU + S y^m y^m \ , \qquad m=1,\dots, n \ ,
\end{equation}
where again the first expression is in terms of $X^I$ and the second one in terms of holomorphic coordinates with $X^0=1$.
As before, we will focus on $M_{\rm v}$ in the following with the discussion for $M_{\rm h}$ being completely analogous.
The K\"ahler potential
is given by
\begin{equation} \label{eq:ex_Kahlerpot2}
 K = - \ln\iu (\bar S - S) - \ln \big(- (T - \bar T)(U - \bar U)- (y^m - \bar y^m)(y^m - \bar y^m)\big) \ ,
\end{equation}
so that the K\"ahler cone condition \eqref{eq:Kahlercone} reads
\begin{equation}
 \Im S ( \Im T \Im U + \Im y^m \Im y^m) > 0 \ .
\end{equation}
In the following we will concentrate on the domain where
\begin{equation}\label{eq:ex_Kahlercone}
 \Im S > 0 \ , \qquad \Im T \Im U + \Im y^m \Im y^m > 0 \ .
\end{equation}

In \cite{Cortes:2011ut} the condition \eqref{eq:cmapMinkN1} was discussed in detail for the example \eqref{eq:ex_prep}. The vector $C^\Lambda$ parametrising the embedding tensor was  defined to be
\begin{equation}\label{eq:ex_CI}\begin{aligned}
C_S = \frac{C^T C^U}{C^0} \ , \qquad C_T = \langle S \rangle C^U \ , \qquad C_U = \langle S \rangle C^T \ , \\
C_m = 2 \langle S \rangle  C^m \ , \qquad C_0 = - \langle S \rangle \frac{C^T C^U}{C^0} \ , \qquad C^S = \langle S \rangle C^0 \ ,
\end{aligned}\end{equation}
with $C^0 \ne 0$.
The $N=1$ vacuum is located at
\begin{equation}
 S = \langle S \rangle \ , \qquad  (T - \frac{C^T}{C^0})(U -
 \frac{C^U}{C^0}) + (y^m - \frac{2C^m}{C^0})\, y^m = 0 \ .
\end{equation}
If $\Im \langle S \rangle > 0 $, condition \eqref{eq:lightlike} gives
\begin{equation}\label{eq:ex_lightlike2}
 \Im \frac{C^T}{C^0}  \Im \frac{C^U}{C^0} = - \frac{C^m \bar C^m}{2|C^0|^2} \ .
\end{equation}
If we take $\Im \frac{C^T}{C^0} > 0$, then one point of the $N=1$ vacuum is given by
\begin{equation}
 T = \frac{C^T}{C^0} \ , \qquad y^m = 0 \ ,
\end{equation}
and therefore an $N=1$ vacuum exists.

Now let us discuss the $N=2$ vacuum. From \eqref{eq:cmapMinkN2} we
obtain two equations that read
\begin{equation} \label{eq:ex_eqN2}\begin{aligned}
(S- \langle S \rangle)\big((T - \frac{C^T}{C^0})(U - \frac{C^U}{C^0}) + y^m (y^m-\frac{2C^m}{C^0}) \big) =&\ 0\ , \\
(\bar S- {\langle S \rangle})\big((\bar T - \frac{C^T}{C^0})(\bar U - \frac{C^U}{C^0}) + \bar y^m (\bar y^m-\frac{2C^m}{C^0})\big) = &\ 0\ .
\end{aligned}\end{equation}
The first one is easily satisfied by $S=\langle S\rangle$. The second one
is then more difficult to solve since \eqref{eq:ex_Kahlercone}
demands $\Im S > 0$.
Here we only display one point of the $N=2$ vacuum to prove that it exists inside the K\"ahler cone. This point is
\begin{equation} \label{eq:ex_solN2}\begin{aligned}
S = & \langle S \rangle \ , \qquad
U =  \Re \frac{C^U}{C^0} + 3 \iu \Im \frac{C^U}{C^0} \ , \\
T = & \Re \frac{C^T}{C^0} + 3 \iu \Im \frac{C^T}{C^0} \ , \qquad
y^m =  2 \iu \Im \frac{C^m}{C^0} \ ,
\end{aligned}\end{equation}
where we set $\Re \frac{C^m}{C^0} =0$. By using
\eqref{eq:ex_lightlike}, one can check that the point
\eqref{eq:ex_solN2} solves \eqref{eq:ex_eqN2} and therefore gives an
$N=2$ solution. Furthermore, \eqref{eq:ex_solN2} lies inside the
K\"ahler cone defined by \eqref{eq:ex_Kahlercone}. Therefore, we have
an $N=1$ and an $N=2$ vacuum in the same $N=2$ gauged
supergravity.

\section*{Acknowledgements}
We would like to thank Zohar Komargodski, Thomas van Riet, Diederik Roest and Bert Vercnocke for useful conversations.
This work was partly supported by the German Science Foundation (DFG) under the
Collaborative Research Center (SFB) 676 ``Particles, Strings and the Early Universe''.
The work of P.S. is supported by the Swiss National Science Foundation.
The work of H.T. is supported by the DSM CEA/Saclay, the ANR grant 08-JCJC-0001-0
and the ERC Starting Independent Researcher Grant 240210 - String-QCD-BH.


\begin{thebibliography}{10}

\bibitem{Hristov:2009uj}
  K.~Hristov, H.~Looyestijn and S.~Vandoren,
  ``Maximally supersymmetric solutions of D=4 N=2 gauged supergravity,''
  JHEP {\bf 0911}, 115 (2009)
  [arXiv:0909.1743 [hep-th]].

\bibitem{deWit:2005ub}
  B.~de Wit, H.~Samtleben and M.~Trigiante,
  ``Magnetic charges in local field theory,''
  JHEP {\bf 0509}, 016 (2005)
  [arXiv:hep-th/0507289].

\bibitem{deWit:2011gk}
  B.~de Wit and M.~van Zalk,
  ``Electric and magnetic charges in N=2 conformal supergravity theories,''
  JHEP {\bf 1110}, 050 (2011)
  [arXiv:1107.3305 [hep-th]].

\bibitem{Huebscher:2010ib}
  M.~Huebscher, T.~Ortin and C.~S.~Shahbazi,
  ``The Tensor Hierarchies of Pure N=2,d=4,5,6 Supergravities,''
  JHEP {\bf 1011}, 130 (2010)
  [arXiv:1006.4457 [hep-th]].

\bibitem{Cecotti:1988qn}
S.~Cecotti, S.~Ferrara and L.~Girardello, ``Geometry of Type II
  Superstrings and the Moduli of Superconformal Field Theories,''  Int. J.
  Mod. Phys. {\bf A4}, 2475 (1989).

\bibitem{Ferrara:1989ik}
  S.~Ferrara and S.~Sabharwal,
  ``Quaternionic Manifolds for Type II Superstring Vacua of Calabi-Yau Spaces,''
  Nucl.\ Phys.\ B {\bf 332}, 317 (1990).

\bibitem{Gauntlett:2004hh}
  J.~P.~Gauntlett, D.~Martelli, J.~F.~Sparks and D.~Waldram,
  ``A New infinite class of Sasaki-Einstein manifolds,''
  Adv.\ Theor.\ Math.\ Phys.\  {\bf 8}, 987 (2006)
  [hep-th/0403038].

\bibitem{Petrini:2009ur}
  M.~Petrini and A.~Zaffaroni,
  ``N=2 solutions of massive type IIA and their Chern-Simons duals,''
  JHEP {\bf 0909}, 107 (2009)
  [arXiv:0904.4915 [hep-th]].

\bibitem{Lust:2009mb}
  D.~Lust and D.~Tsimpis,
  ``New supersymmetric AdS(4) type II vacua,''
  JHEP {\bf 0909}, 098 (2009)
  [arXiv:0906.2561 [hep-th]].

\bibitem{Ferrara:1995gu}
  S.~Ferrara, L.~Girardello and M.~Porrati,
  ``Minimal Higgs branch for the breaking of half of the supersymmetries in N=2 supergravity,''
  Phys.\ Lett.\ B {\bf 366}, 155 (1996)
  [hep-th/9510074].

\bibitem{Ferrara:1995xi}
  S.~Ferrara, L.~Girardello and M.~Porrati,
  ``Spontaneous breaking of N=2 to N=1 in rigid and local supersymmetric theories,''
  Phys.\ Lett.\ B {\bf 376}, 275 (1996)
  [hep-th/9512180].

\bibitem{Fre:1996js}
  P.~Fre, L.~Girardello, I.~Pesando and M.~Trigiante,
  ``Spontaneous N=2 $\to$ N=1 local supersymmetry breaking with surviving compact gauge group,''
  Nucl.\ Phys.\ B {\bf 493}, 231 (1997)
  [hep-th/9607032].

\bibitem{Louis:2009xd}
  J.~Louis, P.~Smyth and H.~Triendl,
  ``Spontaneous N=2 to N=1 Supersymmetry Breaking in Supergravity and Type II String Theory,''
  JHEP {\bf 1002}, 103 (2010)
  [arXiv:0911.5077 [hep-th]].

\bibitem{Louis:2010ui}
  J.~Louis, P.~Smyth and H.~Triendl,
  ``The N=1 Low-Energy Effective Action of Spontaneously Broken N=2 Supergravities,''
  JHEP {\bf 1010}, 017 (2010)
  [arXiv:1008.1214 [hep-th]].

\bibitem{Cortes:2011ut}
  V.~Cortes, J.~Louis, P.~Smyth and H.~Triendl,
  ``On certain K\"ahler quotients of quaternionic K\"ahler manifolds,'' {\rm to appear in} Comm.\ Math.\ Phys.\
  [arXiv:1111.0679 [math.DG]].



\bibitem{Andrianopoli:1996cm}
  L.~Andrianopoli, M.~Bertolini, A.~Ceresole, R.~D'Auria, S.~Ferrara, P.~Fre and T.~Magri,
  ``N=2 supergravity and N=2 superYang-Mills theory on general scalar manifolds: Symplectic covariance, gaugings and the momentum map,''
  J.\ Geom.\ Phys.\  {\bf 23}, 111 (1997)
  [hep-th/9605032].

\bibitem{Michelson:1996pn}
  J.~Michelson,
  ``Compactifications of type IIB strings to four-dimensions with nontrivial classical potential,''
  Nucl.\ Phys.\ B {\bf 495}, 127 (1997)
  [hep-th/9610151].

\bibitem{D'Auria:1990fj}
  R.~D'Auria, S.~Ferrara and P.~Fre,
  ``Special and quaternionic isometries: General couplings in N=2 supergravity and the scalar potential,''
  Nucl.\ Phys.\ B {\bf 359}, 705 (1991).

\bibitem{Gauntlett:2009zw}
  J.~P.~Gauntlett, S.~Kim, O.~Varela and D.~Waldram,
  ``Consistent supersymmetric Kaluza-Klein truncations with massive modes,''
  JHEP {\bf 0904}, 102 (2009)
  [arXiv:0901.0676 [hep-th]].

\bibitem{Gunaydin:1999zx}
  M.~Gunaydin and M.~Zagermann,
  ``The Gauging of five-dimensional, N=2 Maxwell-Einstein supergravity theories coupled to tensor multiplets,''
  Nucl.\ Phys.\ B {\bf 572}, 131 (2000)
  [hep-th/9912027].

\bibitem{Bergshoeff:2004kh}
  E.~Bergshoeff, S.~Cucu, T.~de Wit, J.~Gheerardyn, S.~Vandoren and A.~Van Proeyen,
  ``N = 2 supergravity in five-dimensions revisited,''
  Class.\ Quant.\ Grav.\  {\bf 21}, 3015 (2004)
  [Class.\ Quant.\ Grav.\  {\bf 23}, 7149 (2006)]
  [hep-th/0403045].

\bibitem{Zagermann:2001ki}
  M.~Zagermann,
``The gauging of vector- and tensor-field-coupled five-dimensional N = 2 supergravity,''
  Class.\ Quant.\ Grav.\  {\bf 18} (2001) 3197.

\bibitem{Nishino:1986dc}
  H.~Nishino and E.~Sezgin,
  ``The Complete N=2, D = 6 Supergravity With Matter And Yang-mills Couplings,''
  Nucl.\ Phys.\ B {\bf 278}, 353 (1986).

\bibitem{Riccioni:2001bg}
  F.~Riccioni,
  ``All couplings of minimal six-dimensional supergravity,''
  Nucl.\ Phys.\ B {\bf 605}, 245 (2001)
  [hep-th/0101074].

\bibitem{Polchinski:1995sm}
  J.~Polchinski and A.~Strominger,
  ``New vacua for type II string theory,''
  Phys.\ Lett.\ B {\bf 388}, 736 (1996)
  [hep-th/9510227].

\bibitem{Gurrieri:2002wz}
  S.~Gurrieri, J.~Louis, A.~Micu and D.~Waldram,
  ``Mirror symmetry in generalized Calabi-Yau compactifications,''
  Nucl.\ Phys.\ B {\bf 654}, 61 (2003)
  [hep-th/0211102].

\bibitem{D'Auria:2004tr}
  R.~D'Auria, S.~Ferrara, M.~Trigiante and S.~Vaula,
  ``Gauging the Heisenberg algebra of special quaternionic manifolds,''
  Phys.\ Lett.\ B {\bf 610}, 147 (2005)
  [hep-th/0410290].

\bibitem{Grana:2005ny}
  M.~Grana, J.~Louis and D.~Waldram,
  ``Hitchin functionals in N=2 supergravity,''
  JHEP {\bf 0601}, 008 (2006)
  [hep-th/0505264].

\bibitem{Grana:2006hr}
  M.~Grana, J.~Louis and D.~Waldram,
  ``SU(3) x SU(3) compactification and mirror duals of magnetic fluxes,''
  JHEP {\bf 0704}, 101 (2007)
  [hep-th/0612237].

\bibitem{Cassani:2007pq}
  D.~Cassani and A.~Bilal,
  ``Effective actions and N=1 vacuum conditions from SU(3) x SU(3) compactifications,''
  JHEP {\bf 0709}, 076 (2007)
  [arXiv:0707.3125 [hep-th]].


\bibitem{RoblesLlana:2006ez}
  D.~Robles-Llana, F.~Saueressig and S.~Vandoren,
  ``String loop corrected hypermultiplet moduli spaces,''
  JHEP {\bf 0603}, 081 (2006)
  [hep-th/0602164].

\bibitem{Weinberg:1982id}
  S.~Weinberg,
  ``Does Gravitation Resolve the Ambiguity Among Supersymmetry Vacua?,''
  Phys.\ Rev.\ Lett.\  {\bf 48}, 1776 (1982).

\bibitem{Cvetic:1992st}
  M.~Cvetic, S.~Griffies and S.~-J.~Rey,
  ``Nonperturbative stability of supergravity and superstring vacua,''
  Nucl.\ Phys.\ B {\bf 389}, 3 (1993)
  [hep-th/9206004].

\end{thebibliography}
\end{document}